\documentclass[thmsa,11pt]{article}
\usepackage{sw20lart}
\usepackage{amsmath}
\usepackage{graphicx}
\usepackage{amsfonts}
\usepackage{amssymb}

\begin{document}

\author{Scott Hitchcock\\National Superconducting Cyclotron Laboratory\\Michigan State University, East Lansing, MI 48824-1321 \\NSCL Publication: MSUCL-1135\\E-mail: hitchcock@nscl.msu.edu}
\title{\textbf{Feynman Clocks, Causal Networks, and the Origin of Hierarchical
'Arrows of Time' in Complex Systems. Part I. 'Conjectures' }}
\date{February 4, 2001}
\maketitle
\begin{abstract}
A theory of \textbf{time} as \textbf{information} is \emph{outlined} using new
tools such as \textbf{Feynman Clocks (FC),} \textbf{Collective Excitation
Networks (CENs)}, and \textbf{Sequential Excitation Networks (SENs)}.
\end{abstract}
\tableofcontents

\section{Introduction}

\begin{quotation}
\emph{Should we be prepared to see some day a new structure for the
foundations of physics that does away with time?...Yes, because ''time'' is in
trouble.-John Wheeler \cite{davies}.}

\emph{'Physical time' will emerge as a sort of secondary collective variable
in the network, i.e. being different from the clock time (while being of
course functionally related to it)- Manfred Requardt \cite{cellnets}.}
\end{quotation}

It has been suggested by Julian Barbour that 'time' does not exist
\cite{timeless}. It is the position of this author that 'time' is in fact does
'exist' and is a different 'property' of evolving systems than has been
previously assumed.

Conventional 'time' is \emph{functionally related }to the \emph{signals
created by reconfiguration or 'decoherence'}\cite{zurek}\emph{\ transitions
between the physical states of clocks.} These signals are \emph{detected} by
the conversion of a signal into an excited state in a detector. This process
produces \emph{state information} (e.g. \emph{configuration} observables such
as energy) in the detector connecting it to the source and its signal on an
'arc' between two nodes of a \emph{causal network}.

The detector states can the be sequentially 'clocked' into ordered memory
registers through a process of \textbf{'signal mapping'}.\ The state
information stored in these registers can be used to create a 'time dimension'
by \emph{mapping} the ordered set of memory states onto the real number line.
\emph{Information processing systems} (e.g. quantum computer 'gates', neurons,
the brain etc.) are an integral part of the creation of 'time' as a measure of
the difference between configurations of a system with respect to a 'standard clock'.

Much of the confusion about the \emph{nature of time} is connected with the
\emph{spatialization} of 'time'. The use of 'shift' vectors and 'lapse'
functions \cite{kolb} takes 'explicit' clock time and masks it in an implicit
form of 'distance' between 'universal state configurations'. These are
specified by the distribution of matter and energy in 'space' (vacuum).

The 'problem of time' is primarily about the emergence of macroscopic
irreversibility (e.g. entropy) from reversible microscopic symmetries (the
'$T$'' of CPT invariance). The \emph{quantum arrow of time} is defined by the
\emph{irreversible} 'decay of a discrete state resonantly coupled to a
continuum of final states' \cite{Diu} observed in various nuclear and atomic
processes.\ We will see that apparent 'time' reversibility and irreversibility
are compatible and necessary aspects of quantum systems. The '\emph{Program of
Decoherence}', the entanglement of quantum states, and the emergence and decay
of novel collective excitations provide tools \cite{naqw}, \cite{qdae},
\cite{hologram} for understanding the common roots of all arrows of time for
unstable configurations of hierarchically scaled clusters of matter in an
evolving universe. This scaling leads to 'classical' or macroscopic aspects of
reality that are in fact special cases of a 'quantum universe' \cite{dacwqt}.

We begin with the premise that \textbf{time is a number created by the
processing of information}.\ Scalar division of this information by Planck's
constant creates a real number with units of 'time' (seconds). The information
(in this context has units of 'energy') is propagated by \textbf{signals}
between \textbf{Feynman Clocks (FCs) }to\textbf{\ Feynman Detectors (FDs)}.
The FD is the signal absorption mode of the FC and unless indicated, 'FC' will
be used to represent these two modes of a single system. The conversion of
this state information into numbers provides the basis for building the 'time
dimension' of space-time by measurement of the differences between two
numbers. The creation of these numbers from the detection events is
accompanied by the loss of information about the states of the signals and
clocks. The increase in information entropy will be important in understanding
the 'lifetimes' of local and distributed information structures in quantum computation.

The \textbf{special theory of time} describes how the fundamental quantum
mechanisms involved in the reconfigurations of unstable systems generate
information states defining an irreversible 'quantum arrows of time'. The
\textbf{general theory of time} describes how the quantum arrow of time can be
used to define macroscopic arrows of time associated with transfer of state
information in complex systems and causal networks \cite{hitchcock}. It is the
author's position that unified or comprehensive 'theories of everything,
everywhere, at anytime' require a \emph{deep} understanding of 'time'.
Conventional ideas of 'time' have been the implicit and explicit source of
many contradictions and paradoxes in physical theories.

The purpose of this paper is to examine the premise that all temporal
processes in macroscopic complex systems can be understood as being generated
by a microscopic \textbf{irreversible quantum arrow of time.} The
\emph{correspondence} between the various separate biological, cosmological,
psychological, radiative, and thermodynamic '\emph{arrows of time'}
\cite{zeh}, \cite{cove}, \cite{schulman1}, \cite{rirnr} is achieved with
\emph{causal networks} built up hierarchically from the quantum arrow of time.
\emph{Preliminary} and \emph{speculative} ideas and tools are presented in
order to see if a 'deeper' descriptive and computational 'language' of 'time'
as 'information' is possible.

\section{An Approach to 'Time'}

The description of 'time' explored in this paper is built on the following conjectures:

\begin{conjecture}
'\textbf{Time}' is a \textbf{form} of '\textbf{information'}.
\end{conjecture}

\begin{conjecture}
\textbf{Information} is created by \textbf{reconfigurations} of unstable
states of systems.
\end{conjecture}

\begin{conjecture}
\textbf{Reconfigurations} of systems produce or process '\textbf{signals}'.
\end{conjecture}

\begin{conjecture}
\textbf{Signals} connect \textbf{clocks} to \textbf{detectors}. A
\textbf{Feynman Clock (FC) or 'gate'} is a generalization of a \textbf{quantum
clock} \cite{Peres} with multiple signal input and output 'processing'
capabilities. Signals are essential for the creation of causal networks.
Signals are 'quantum' in nature but may appear 'classical' as a result of
their collective scale, statistical entanglement, or intensity. They may be
identified with a transition in a macroscopic Plateau of Complexity or 'POC'
(see below).
\end{conjecture}

\begin{conjecture}
The \textbf{Quantum Arrow of Time (QAT)} is a \textbf{pointer. }It is a
function mapping the \textbf{irreversible} transition from an excited or
unstable configuration to a coupled 'stable' one in a FC system. The
'direction' and 'magnitude' of these 'arrows' are specific to the quantum
system being observed. The general QAT is really a statement about unstable
states of system coupled to one or more states from within a set of all
possible reconfigurations of that system. These pointers are
'\textbf{information vectors}' in an '\textbf{information space}'
\cite{infospace}. The state information transfer between clocks and detectors
by the signals can be mapped by these vectors in the information space. All
QATs for various localized systems in the current universe are 'traceable'
back to the fundamental QAT mapping the decoherence and decay of the initial
excited state of a FC-Universe at the beginning of the Big Bang.

\begin{conjecture}
The \textbf{creation} of an \textbf{unstable state} in a system by the
\textbf{detection} of one or more simultaneous or staggered \textbf{signals}
is the '\textbf{detector' mode }of a clock.
\end{conjecture}

\begin{conjecture}
The '\textbf{decay}' or '\textbf{decoherence}' of this state with the emission
of one or more \textbf{signals} is the '\textbf{clock' mode} of the system.
\end{conjecture}

\begin{remark}
\textbf{Note} that the term '\textbf{clock' or FC} will be used to represent
any system that has \textbf{both} detector and clock modes unless otherwise
noted. A \textbf{Feynman Detector (FD) }is the \textbf{input \ }%
or\textbf{\ detection mode} of a Feynman Clock.
\end{remark}
\end{conjecture}

\begin{conjecture}
\textbf{Causal Networks} are \textbf{built} from locally \textbf{connected}
sets of clock (and detector) \textbf{nodes.} These networks map the sequential
progression of information or signal flow from node to node. They may also act
as '\textbf{temporal interferometers}' acting on or creating \textbf{entangled
signal} states \cite{qcent}.
\end{conjecture}

\begin{conjecture}
Causal networks of signal linked FCs processing signals \textbf{sequentially}
are \textbf{Sequential Excitation Networks (SENs).} These systems treat the
signals between nodes as distinct 'classical' objects which are decoupled from
the clocks and detectors along their trajectories. We will see that the
distinction between quantum and classical objects is conceptually artificial.
It arises when systems exhibit collective properties that can be described
without direct reference to the complex quantum causal networks underlying
them. 'Classical' methods are clearly appropriate for the application of
classical physics and engineering principles in macroscopic systems where
their possible disruption by underlying quantum CEs is negligible. .
\end{conjecture}

\begin{conjecture}
\textbf{FCs} can be ''\textbf{synchronized}'' by \textbf{entangled} signals by
generalizing the ''\textbf{Quantum Clock Synchronization Scheme}''
\cite{qcent}. The FCs may be separated by 'classical' distances but act as a
single quantum system. The spatial 'distance' between two coupled or entangled
nodes is a 'weak' measure of the quantum or classical nature of the combined
system. The 'distance' which distinguishes the two nodes may be more properly
'measured' by their\textbf{\ degree of entanglement} \cite{reqit}.
\end{conjecture}

\begin{conjecture}
\textbf{Collective Excitations (CEs)} (e.g. phonons) of \textbf{synchronized}
sets of Feynman clocks in causal networks result from reconfigurations of
excited state of the whole system. These CEs occur as novel behaviors of these
\textbf{Collective Excitation Networks (CENs)}.
\end{conjecture}

\begin{conjecture}
\textbf{Plateaus of Complexity (POCs)} are the \textbf{physical}
\textbf{states} of complex systems (CENs) that support \textbf{CEs }and
the\textbf{\ collective signals} that are\textbf{\ }detected\textbf{\ }%
and\textbf{\ }emitted by the collective state of the components acting
together as a single 'clock'. Signals are essential for the creation of causal
networks. Signals are always quantum in nature but may appear classical as a
result of the identification with a macroscopic POC.

\begin{remark}
These CEs are perturbations of POC $n$-body states. They may be multipole
oscillations of nuclei shifted energetically away from the mean POC energy
states associated with 'stable' configurations. These stable nuclei occur at
''\textbf{magic}'' total nucleon numbers given by the proton and neutron sum
$A$, where $A=2,8,20,28,50,82,$or $126$ \ calculated using the 'Shell Model'
of nuclear structure\cite{heyde}.
\end{remark}
\end{conjecture}

\begin{conjecture}
The \textbf{spatial direction} of the \textbf{flow of information} via signals
in networks defines '\textbf{arrows of time'} specific to the \textbf{signals}
and their \textbf{information content.}
\end{conjecture}

\begin{conjecture}
The emergence and decay of \textbf{CEs }in \textbf{POCs }generate signals that
can be used to create \textbf{hierarchical 'arrows of time'} associated with a
system at a given level of complexity.
\end{conjecture}

\begin{conjecture}
Complex systems \textbf{composed} of other complex systems can generate
\textbf{POCs within POCs}. POCs can act as \textbf{clocks} forming
\textbf{networks of POCs} which in turn can support new CEs. This
'\textbf{nesting}' of clocks and signals provides the basis for scaling from
quantum clocks to neural networks. The signals between such systems that are
\emph{not acting collectively} form hierarchical \textbf{SENs}.
\end{conjecture}

\begin{conjecture}
\textbf{All 'arrows of time}' defined for various POCs are \textbf{traceable}
back to fundamental QATs.
\end{conjecture}

\begin{conjecture}
The '\textbf{direction}' and '\textbf{dimension}' of 'arrows of time' are
created through a process of \textbf{Signal Mapping }in which the detection of
non-simultaneous signals is \textbf{causally ordered} in a \textbf{memory
register} by \textbf{coupling} the detector states to signals generated by an
\textbf{internal} or \textbf{standard} \textbf{clock \cite{rirnr}}. The set of
ordered states can then be transformed by a computer (e.g. neural network) to
the set real numbers for the construction of a 'timeline'. This timeline can
then be converted into the \emph{spatialized} 'time' parameter of relativistic 'space-time'.
\end{conjecture}

\begin{conjecture}
\textbf{QATs are irreversible} since they are defined as originating from
unstable configurations.
\end{conjecture}

\begin{conjecture}
'Time' or more properly \textbf{Configuration State Information} is
\textbf{reversible} or \textbf{restorable \cite{rirnr}\ }\emph{only if an
unstable state of a system can be recreated} by work (injection of the
wave-function information specific to the desired state) done on the system by
an external 'agent'(signal). This work may be done as a quantum computation on
\textbf{reconfiguration signals} by the \textbf{FD 'gates'}. This can create a
\textbf{local} 'reversal of entropy' leading to an \textbf{excited}
configuration of matter with the same properties as the atemporal 'previous
'\ state of the system. The creation or re-creation of an unstable state of a
system requires input information from the systems environment and is not
associated with an internal QAT but with the signals QAT..
\end{conjecture}

\section{The Quantum Arrow of Time (QAT)}

The decay of \emph{ensembles} of identical radioactive nuclei can be described
in the 'bulk' or statistical perspective with the '\textbf{exponential decay
law}' \cite{ourfriend}. We will see that the system as a whole can be thought
of as decaying form the initial collective\ excitation of a 'network' of
acausal but 'connected' set of quantum systems. The 'decay law' in the
ensemble paradigm is;%

\begin{equation}
N_{\tau}=N_{0}e^{-k\tau}%
\end{equation}

where $N_{0}$ is the number of identical radioactive 'clocks' that we start
with, $k$ is the 'decay constant' specific to the type of clock and represents
the magnitude of the 'instability' of the system, $\tau$ is the 'elapsed time'
for the transition from an \emph{initial system configuration state} of
$N_{0}$ clocks to a \emph{reconfigured state }of the system where $N_{\tau}$
clocks remain at 'time' $\tau$. Since the unstable nuclei of this system are
quantum clocks, we can view this equation as the recipe for a 'collective' or
'statistical' clock built from an ensemble of quantum clocks. Solving for
'time' we have;%

\begin{equation}
\tau_{edl}=\frac{1}{k}\left(  \ln\frac{N_{0}}{N_{\tau}}\right)
\end{equation}

which can be thought of as the \emph{transformation} of '\emph{reconfiguration
information}' on the right into the '\textbf{lifetime}' of the transition from
the unstable state of the initial system to a 'more stable' state\ with
$N_{\tau}$\ clocks. The 'detection' of the$\ N_{\tau}$\ 'signal' results in a
real number created by the \emph{dimensional conversion} factor $k^{-1}$. This
number is interpreted as an 'event time' for the state of a statistical clock
built from many quantum clocks.

First we observe that the decays of radioactive nuclei, excited electronic
states of atoms through 'autoionization' (emitting a 'free' electron) or
photon emission are described with '\textbf{time-independent}' perturbation theory;

\begin{quotation}
''For example, a system, initially in a discrete state, can split, under the
effect of an internal coupling (described, consequently, by a time-independent
Hamiltonian $W$), into two distinct parts whose energies (kinetic in the case
of material particles and electromagnetic in the case of photons) can have,
theoretically, any value; this gives the set of final states a continuous
nature...We can also cite the \emph{spontaneous emission }of a photon by an
excited atomic (or nuclear) state: the interaction of the atom with the
quantized electromagnetic field couples the discrete initial state (the
excited atom in the absence of photons) with a continuum of final states (the
atom in a lower state in the presence of a photon of arbitrary direction,
polarization and energy).''\cite{Diu}
\end{quotation}

\ These decay modes are not restricted to atoms and nuclei. We will see that
these \emph{quantum clocks} are \emph{time-independent irreversible systems}
that can be created in space by \emph{apparently} 'time' reversible particle
collisions. \emph{The key to the irreversibility in quantum systems is the
creation of an \textbf{unstable} configuration of matter and energy in space}.
This is the '\emph{first cause}' for decay. The decay of an unstable state
creates a 'local' arrow of time pointing to more stable states for the system.
\emph{Instability} is a measure of the \emph{geometric asymmetry} of the
mass-energy distribution of the 'components' as they are driven to 'more
stable' configurations by the fundamental interactions (forces) between each other.

The apparent 'reversibility' of clocks or any other complex system is an
phenomena created by the interaction of spatially distinct signals and
detectors. Reversibility is a collective property of a composite system formed
with 'free' signals, the vacuum, and quantum detectors and clocks. 'Entropy'
is a convenient mapping tool or system pointer indicating the direction of
evolution of the total system\ while preserving the reversibility of detectors
and the irreversibility of clocks.

\section{Collective Excitations}

The key to hierarchical systems of quantum systems acting as 'classical'
objects is the concept of \textbf{collective} \textbf{excitations (CEs)} of
\textbf{quasiparticles }(also called 'elementary excitations') \cite{Mattuck},
\cite{zago}, \cite{heyde}. \textbf{Phonons}, \textbf{excitons}, and
\textbf{plasmons} are examples of CEs that exhibit mesoscopic system behaviors
but are still quantum phenomena \cite{zim}, \cite{ssp}. Superconductivity
represents an important quantum macroscopic behavior in the CE model.

If spatially extended quantum CEs can be shown to exist in complex 'classical'
systems, then CE emergence and decay can define the 'lifetime' of the
reconfiguration process involved with transitions between specific states.
Collective excitations represent new properties of $n$-body aggregates of
matter that are not a mere sum of the individual properties of the individual
components. The emergence of these novel behaviors creates the opportunity for
new 'signals' to be emitted and absorbed as resonances of the new energy
eigenstates. The state information transported by these signals is also new.
These signals allow identification of the transitions between states in the
spectrum of CE states of a complex system. These phonon or phonon-like signals
can link other CE systems in hierarchical causal networks.

Recent work on the synchronization of quantum clocks provides a model for CEs
as entangled states in widely separated systems through a ''\emph{quantum
clock synchronization scheme}'' \cite{qcent}. This model can be expanded for
Feynman Clock Synchronization over 'classical' distances where the FCs are
virtual clocks (entangled 'time' independent \emph{signals}) until 'measured'
or decohered from an atemporal global CE state into 'actual' FC states of the
nodes in a causal network. These synchronized nodes create a CEN without the
exchange of 'timing information'. Evidence of CEs over great distances is
found in photon entanglement experiments.

Experimental observation of two 'energy-time' entangled photons separated by
more than 10 Kilometers \cite{genFC} provides an example of the \emph{decay of
a collective excitation} of a vary large spatially extensive quantum system
\emph{if} \ we look at the entire experimental setup as a 'SEN' system from
the 'Geneva FC' to the Bellevue/Bernex 'CEN'. The 'Geneva FC' produces two
'coherent' photon signals that traverse large distances on separate fiber
optic paths (8.1 and 9.3 km). The 'transit lifetimes' of the signals are
functions of the velocity of the signals in the medium and their distances to
the FDs in the Bellevue/Bernex CEN. Signal mapping of the FD/FC detection
events in the CEN via a 'clocked' memory system linking the two 'node' leads
to causal ordering. The entangled photons remained 'correlated' even though
separated by 10.9 kilometers, upon their detection 'decohere' with the
production of 'classical' information (i.e. the emission of 'signals' or the
creation of 'states' in memories) upon measurement.

The existence of spatially extended quantum states in networks depends on the
entanglement of the states of the components. Entanglement allows CEs to
emerge and decay. The lifetimes of these states is controlled by
environmentally induced decoherence. \emph{Decoherence lifetimes} can be
extended by 'self-measurement' or 'feedback' with the CE or environment. This
allows for the existence of macroscopic quantum states of networks. The
lifetimes can also be shortened by decohering signals causing 'feedforward' of
the evolution of the system. If the interaction extends the lifetime of the
state then entropy in minimized. If the lifetime is shortened then entropy is
maximized. Information loss is a measure of the entropy of the system. It is
lost via emitted signals to the environment.

Quantum entanglement of the states of many components of a network can then
'define' the collective excitation as the \emph{resultant state} of the
coupled interactions of all the nodes. CEs may also interact with each other.
This may lead to entanglement of various states of a plateau of complexity
creating a higher order CE. Nesting of sets of entangled states within causal
networks can lead to an entangled state composed of entangled states. This may
provide a basis for the existence of spatially extended complex higher order
CEN quantum states over 'classically' separated nodes of a causal network.

The emergence of classical systems as collective effects of networks of
quantum systems through a process of collective excitation state signal
production and the 'decoherence' of quantum superpositions of states into
'classical' signals and systems calls for a description in which the lifetimes
of unstable states of systems of all sizes 'correspond' in a logical way to
the reconfiguration processes of their subsystems. For an 'open' system (e.g.
an atoms electron in an excited state) the decoherence can be induced its
coupling to its 'environment' (e.g. the 'vacuum' plus QED
'self-interaction').\ The definition of the systems' environment depend of the
specific property or state of the system is interacting with the 'external'
space in which it is embedded. For example the 'vacuum' may not be the
relevant environment for the chemically driven metabolic activities of a cell
although it 'exists' between the chemicals. However it is an essential
'environment' for virtual particle production and decay and the 'Casimir'
forces of attraction between two flat parallel plates in it \cite{ipp}. The
initial state of the universe can be considered to be self-contained 'closed'
system with no 'external' environment. It can however have an 'environment'
composed of primordial density perturbations from CE 'phonons'. The phonon
states may have been frozen out as hierarchical clusters of matter or 'caused'
the decoherence of the initial state of the Universe driving inflation and the
Big Bang.

If we look at the boundary conditions (e.g. gravitation) defining a closed
\emph{system} of mass-energy as it's \emph{apparatus} then the unstable
initial state of the universe is due to the \emph{collective excitation
'environment'} of the system plus its boundary conditions. The excited state
may be the result of 'constructive' interference of some or all of the
possible reconfiguration states.

In collective modes of $n$-body nuclei \cite{heyde}, the CE 'environment' is
the phonon field with a characteristic multi-phonon spectrum. The 'boundary
condition' acting as an apparatus 'measuring' the nuclear configuration state
of the system is the 'surface tension' due to the binding energy of the strong
interaction between nucleons in the nucleus. The phonon resonances of the
nucleus and its 'surface' represent a prototypical collective excitation that
emerges at a plateaus of complexity for this CE. 'Giant resonance' collective
excitations of nuclei emerge from the coherent states of the nucleons. The
resulting decay of the resonance by decoherence of the CE is caused by its
coupling to the non-coherent modes of motion for the nucleus resulting in the
'damping' of the collective motion. If the CE has enough energy to exceed the
stationary state equilibrium energy, then the system 'decays' irreversibly
into a new configuration.

The interaction of the CE with the internal configuration states of the system
is a time independent '\emph{self-measurement}' \cite{tesm} which causes the
system to 'decay' or decohere irreversibly. The CE acts as the 'environment'
coupling the present configuration to all the possible 'future'
reconfiguration states. In the case of the universe the difference between the
non-stationary 'closed' initial state and the 'open' expanding system is the
creation of the 'vacuum' as a decay product. The \textbf{expansion }the
universe is now a \emph{collective excitation} of the mass-energy plus
gravitation system.

The CEs of systems may act as measurements on the internal states by the
surface environment. This surface represents a plateau of complexity for these
systems. These plateaus have collective behaviors including irreversible
transitions to new configurations of matter and energy in expanding space. One
can artificially ascribe scaled arrows of time for these plateaus. These
system dependent arrows are derived from the quantum arrow of time. They
'\emph{correspond}' to the quantum arrow through the collective excitations
and behaviors of the networks of clocks and signals throughout the
hierarchical clusters of information processing subsystems.

\section{Signals}

A signal is any 'system' that conveys information from one system (e.g. FC) to
another (e.g. FD). The creation of a detector state from a signal state is the
end process of information transfer originating in a spatially distinct FC.
The state information transfer causes the reconfiguration of the detection
system resulting in an unstable state of 'excess' information.

The signal path length, $d$, combined with this 'derived' velocity, $v$, (at
this point assumed constant) can be used to find the '\emph{lifetime}' of the
signal from its creation by an FC to its annihilation by a FD. This 'lifetime'
is the classical transit 'time' of the signal given by the macroscopic or
classical relation:%

\begin{equation}
\tau_{signal}=\frac{d}{v}%
\end{equation}

For any 'classical' arbitrary trajectory of a signal in space with a
non-constant velocity function of 3-space position $\mathbf{\vec{r}}$, the
classical velocity function $\mathbf{\vec{V}}(\mathbf{\vec{r}})$, and the
differential signal direction, $d\mathbf{\vec{r}}$, are dependent on a
fundamental interaction of the signal with the medium/environment through
which it passes. The net 'lifetime' of the signal is found by integrating over
the path from source clock position, $C(\mathbf{\vec{r}}_{0})$, to detector
position, $D(\mathbf{\vec{r}}_{l})$. The 'classical' lifetime of the signal is then:%

\begin{equation}
\tau_{signal}=\int\limits_{C(\mathbf{\vec{r}}_{0})}^{D(\mathbf{\vec{r}}_{l}%
)}\left(  \mathbf{\vec{V}}(\mathbf{\vec{r}})\right)  ^{-1}\cdot d\mathbf{\vec
{r}}%
\end{equation}

The 'classical' and 'quantum' lifetimes of signals overlap for cases. The
first case is when the signal trajectory path length is of the order of
collective excitation modes of the system and is bound to the system. This
means that the 'signal' does not propagate 'freely' as in the case of photons
created by decay processes inside a star.

It becomes a 'classical signal' if it propagates 'freely' in space. The
decoupled photon escaping the surface of a star (subject to gravitational
redshift effects) carries state information about its 'last' source in a
networks of sources. A photon can be a quantum or classical signal depending
on the environment in which it propagates. This is really an artificial
separation that is intended to illustrate the subtle nature of the
correspondence principle connecting the quantum description of signals with
the 'classical' electromagnetic wave formalism.

The lifetime of a signal from emission (excited state), propagation
(decoherence or decay lifetime), to its detection ('ground state') can be
viewed as the decay of a single collective excitation of a s-FC system
composed of original source FC, the vacuum or other signal medium, and the FD.
The FC and FD at either ends of the signal trajectory 'bound' the s-FC. The
'quantum lifetime' of the 'signal' or its \emph{equivalent} s-FC system is:%

\begin{equation}
\mathbf{\tau}_{C\rightarrow D}=\frac{\hbar}{\Gamma_{C,D}}=\tau_{signal}%
\end{equation}

where $\Gamma_{C,D}$ is the decay reconfiguration information in the form of
the natural width of the resonance (excitation state).

\section{Feynman Clocks (FCs)}

Feynman diagrams are the source of Feynman clocks created by transforming the
'time' component (dimension) of the incoming and outgoing signals into the
state information content of those signals. The interaction (collision or
scattering) of the incoming signals creates a Feynman clock for the case in
which there was no pre-existing matter in that volume of space. For the case
of a 'target' interacting with incoming signals, the system composed of
absorbed or scattered signals and the target form a Feynman detector in that
volume of space. The target 'detects' the signals in the process of
interaction with them in which new states of the composite system are created.
If this system is unstable, then the Feynman detector mode of the target has
become a Feynman clock. Generally the incoming particles create a clock where
there was no clock before. FCs may be 'open' or 'closed' in relation to the
incoming and outgoing signal trajectories.

For incoming signals whose total momentum is;%

\begin{equation}
\mathbf{p}_{0}=%
{\textstyle\sum\limits_{i=1}^{m}}
\mathbf{p}_{i}%
\end{equation}

resulting in the creation of outgoing signals whose total momentum is;%

\begin{equation}
\mathbf{q}_{0}=%
{\textstyle\sum\limits_{j=1}^{n}}
\mathbf{q}_{j}%
\end{equation}

a 'transient' clock system is created through reconfigurations of the matter
and energy in the signals via the strong, electromagnetic, weak, and
gravitational fundamental interactions (indexed by $I=s,em,w,g$ respectively).
The \textbf{net }Feynman clock 'lifetime' from the system state created by the
interacting incoming signals (FD mode) through the 'decay' process (internal
'decoherence' mode collective excitation state decay) to the state in which
the outgoing decoupled signals are emitted (FC mode) is given by;%

\begin{align}
\mathbf{\tau}_{FC_{net}}  &  =\frac{\hbar}{\Gamma_{FC_{net}}}=\tfrac{\hbar}{%
{\displaystyle\idotsint\limits_{q_{n}\cdots q_{1}}}
\left[  \tfrac{V^{n+1}}{\left(  2\pi\right)  ^{3n+4}}\mathbf{P\cdot}\left|
\mathbf{M}_{I}\right|  ^{2}\delta_{4}\left(  \mathbf{p}_{0}-\mathbf{q}%
_{0}\right)  \right]  dq_{1}dq_{2}\cdots dq_{n}}\\
&  =\tfrac{\hbar}{%
{\displaystyle\idotsint\limits_{q_{n}\cdots q_{1}}}
\left[  \tfrac{V^{n+1}}{\left(  2\pi\right)  ^{3n+4}}\mathbf{P\cdot}\left|
\mathbf{M}_{I}\right|  ^{2}\delta_{4}\left(
{\textstyle\sum\limits_{i=1}^{m}}
\mathbf{p}_{i}-%
{\textstyle\sum\limits_{j=1}^{n}}
\mathbf{q}_{j}\right)  \right]  dq_{1}dq_{2}\cdots dq_{n}}%
\end{align}

If there is no reconfiguration of the incoming signals and target (if any) in
this region of space, then a clock has not been 'created' and the reduced
fundamental interaction matrix element $M_{I}$ (Note: equal to the $S$-matrix
(the 'scattering' matrix) except for the $\delta$-function for overall
energy-momentum conservation) \cite{Veltman} is zero:%

\begin{equation}
\mathbf{M}_{I}=0
\end{equation}

The above equations for the Feynman diagram method for FD/FC 'lifetimes'
represent the creation of 'lifetime' information from a scattering process
that in general is very difficult to compute for complex systems. The idea
here is that a 'collective excitation system' is created by the incoming
signals leading to an irreversible decay with the production of outgoing
signals. The transformation of the incoming signals by collisional
'processing' in a target 'gate' creates new information in the form of the
novel emergent signal states.

\section{Collective Excitation Networks (CENs)}

Collective behaviors of systems composed of discrete but connected components
need to be characterized in order to understand how 'arrows of time' emerge in
complex systems. The concept of 'collective excitations' in the many-body
problem \cite{Mattuck} and in phonon behavior in solids \cite{zim}, \cite{ssp}
provides the basis for modeling reconfigurations of states in causal networks
that represent 'plateaus of complexity' such as the lifetimes of phase in cell
reproduction. When a set of subsystems in a complex system are 'wired'
together in a network, they can act as a coherent superposition of states
capable of supporting new excited states of that network (possibly within
other networks). These collective states have finite lifetimes and decay with
the production of 'signals' (e.g. phonons, solitons, plasmons, 'sound waves', etc.).

The first level of complexity emerges when sets of \emph{coupled }Feynman
clocks act collectively as a single system with new system energy eigenstates
(e.g. molecular spectra) whose unstable excitation modes decay with finite
lifetimes. This system is a\textbf{\ Collective Excitation Network} or
\textbf{CEN}. These CENs can support new \emph{collective excitation states
and signals.} They can also act as 'gates', memories, or registers creating
and processing signals (information) when embedded in larger networks. This
process of 'nesting' of subsystems with collective excitation states provides
a means for deriving various hierarchical 'arrows of time' connected with
plateaus of complexity.

A CEN composed of coupled FCs. Below is the representation of the CEN node for
mapping information flow in causal networks. The next level of temporal
complexity arises when sets of individual Feynman clocks and CENs interact to
form CEN 'circuits' with \emph{multiple inputs and outputs} as in the case of
generalizing a simple quantum clock into a Feynman clock. These 'integrated'
circuits or arrays act as a single unit with collective states unique to the
subsystem as a whole.

The lifetime of a CEN is given by:%

\begin{equation}
\mathbf{\tau}_{CEN}=\frac{\hbar}{\mathbf{\Gamma}_{CEN}}%
\end{equation}

\section{Sequential Excitation Networks (SENs)}

SENs are sets of FCs and CENs processing signals in series.

The 'lifetime' of a SEN is the signal processing 'lifetime' marked by an
incoming input 'signal', $S_{IN}$, to creation of an output signal, $S_{out}$
for a set of FCs, CENs, sub-SENs and the signals between them is the sum of
the lifetimes along the computational path from input to output of each of the
'gates' and signals. The SEN 'lifetime' for this process is given by:%

\begin{equation}
\mathbf{\tau}_{nsum}\equiv\sum\limits_{j}(\tau_{FC_{j}}+\tau_{S_{j}})
\end{equation}
where the CENs and sub-SENs are treated as FCs (note: the FC 'lifetime',
$\tau_{FC_{j}}$, as used here is a sum of the FD lifetime, decoherence
lifetime if any, and the quantum (Feynman) clock decay lifetime for that 'node'.

\textbf{Feedback, feedforward }and\textbf{\ cyclical flow of signals
(information)} is also possible in the SEN. This provides a mechanism for the
resetting of unstable configurations necessary for quantum computational
algorithms. It also provides for adaptive behavior in relatively closed
systems like cells. These 'control' mechanisms can be realized by defining
signal trajectories or 'circuits' connecting various nodes into hybrid linear
and cyclical causal networks. All of the combinatorial possibilities for
'connecting' systems and subsystems together by signal loops provide a means
for modelling complex self-adjusting or adaptive behaviors in which the
continual transformations of the local states or their relative network
configurations of FD/FC, CEN, and SEN nodes produce different computational
'lifetimes' for the information 'currents' propagating through them.

When systems of FCs, CENs, and their signals act as distinct nodes and arcs in
a causally ordered network, the transfer of configuration information through
the system can be 'sequential' from a set of inputs to a set of outputs. If
the entire '\emph{heterogenous}' system acts like a 'black box' clock then
details of the internal processes are embedded in a \textbf{Sequential
Excitation Network }or\textbf{\ SEN }of FCs and CENs.

\section{Plateaus of Complexity (POCs)}

As we have seen above collective excitations are the markers for new levels of
complexity in hierarchically connected systems. Solitons represent 'classical'
wave packet signals in macroscopic scale systems. Their origins are found in
the plateaus of complexity of the subsystems from which they are composed.
Since CEs are the result of the superposition of \emph{quantum states
resulting in another quantum state,} classical states emerge as the result of
the interaction of this system with an \emph{environment}. Plateaus of
complexity are the interface between the quantum properties of the system and
its environment. This is how quantum systems in CENs and SENs can create
'classical' signals and behaviors as a result of the environmental measurement
by an observing system in which it is embedded. The environmental component
makes the quantum system 'open' to classical signal production. If the
environment is the boundary condition on the quantum system it may be 'closed'
but still act like an open system which can decohere (e.g. decay of FC mode of
the initial state of the universe in Big Bang scenarios).

\textbf{Feedback, feedforward }and\textbf{\ cyclical flow of signals
(information)} is also possible in the SEN. This provides a mechanism for the
resetting of unstable configurations necessary for quantum computational
algorithms. It also provides for adaptive behavior in relatively closed
systems like cells. These 'control' mechanisms can be realized by defining
signal trajectories or 'circuits' connecting various nodes into hybrid linear
and cyclical causal networks. All of the combinatorial possibilities for
'connecting' systems and subsystems together by signal loops provide a means
for modelling complex self-adjusting or adaptive behaviors in which the
continual transformations of the local states or their relative network
configurations of FD/FC, CEN, and SEN nodes produce different computational
'lifetimes' for the information 'currents' propagating through them.

\section{Signal Mapping}

Signal mapping is the process by which signals carrying state information are
detected and their 'information content' (induced state in detector) put into
ordered sets with respect to a standard or internal clock. This involves
creating states in a 'memory' so that their causal relation to other events
can be 'labeled' and interpreted. 'Time' emerges as the functional value of
the energy eigenstates in the detectors as information 'bits' assigned to a
detected signal from an 'event' (FC created signal) in 3-space (possibly
$n$-space at the Planck scale for higher dimensional quantum modes of
'strings' etc.).

\subsection{Temporal 'Unification' of the Fundamental Interactions}

For a FC (or CEN in each of the following) reconfigured by the strong
interaction we have a decay lifetime $\mathbf{\tau}_{U}$ :%

\begin{equation}
\mathbf{\tau}_{U}=\alpha\mathbf{\tau}_{strong}=\frac{\hbar}{\mathbf{\Gamma
}_{strong}}%
\end{equation}

For a FC system driven by the weak interaction (or 'electroweak') we have:%

\begin{equation}
\mathbf{\tau}_{U}=\beta\mathbf{\tau}_{weak}=\frac{\hbar}{\mathbf{\Gamma
}_{weak}}%
\end{equation}

For a FC system driven by the electromagnetic interaction (QED photon/electron
processes) we have:%

\begin{equation}
\mathbf{\tau}_{U}=\delta\mathbf{\tau}_{em}=\frac{\hbar}{\mathbf{\Gamma}_{em}}%
\end{equation}

and for a gravitational FC system we have:%

\begin{equation}
\mathbf{\tau}_{U}=\epsilon\mathbf{\tau}_{grav}=\frac{\hbar}{\mathbf{\Gamma
}_{grav}}%
\end{equation}

where the lifetimes are related by real scalar constants $\alpha$, $\beta$,
$\delta$, and $\epsilon$. The unified 'lifetime', $\mathbf{\tau}_{U}$ is then:%

\begin{equation}
\mathbf{\tau}_{U}=\alpha\mathbf{\tau}_{strong}=\beta\mathbf{\tau}%
_{weak}=\delta\mathbf{\tau}_{em}=\epsilon\mathbf{\tau}_{grav}%
\end{equation}

These four prototypical systems are reconfigured by different forces but their
signals provide a rather obvious and perhaps trivial way of establishing an ad
hoc unification of the fundamental interactions of matter in an
\emph{information space} \cite{infospace}. The key to this type of unification
is recognizing the dimensional equivalence of the 'lifetimes' and therefore
the source 'information' common to all the fundamental interactions and that
the systems can act like 'clocks' producing signals that carry information to detectors.

Signals generated in the decay processes above carry state information to
detection systems where the signal generating events can be 'measured' with
respect to each other as functions of 'arrival times', signal spectra energy
distributions, and spatial directions. This process of signal mapping by an
observing system creates the 'times' in the ordered sets of sequential events.
The ordering is with respect to an internal or external standard 'cyclical' FC
system. At this point the \emph{differences} in the order of the detected
signal states can be used to create the 'difference times' or secondary 'event
times' used in the 'coordinates' and time dimension of the space-time of
special and general relativity.

At a subtle level it is the \emph{information} (e.g. specific energy states
associated with 'signals') flow between these systems that may ultimately
provide a context for working 'backward' from collective features of systems
to the unification of physical laws in the microscopic domain of particle
physics. This represents a rather obvious and perhaps trivial way of viewing
the unification of the fundamental interactions of matter from the information
and 'lifetime' frame of reference. The key to unification may be seen in the
'lifetime' or information terms common to all theses interactions. This occurs
in the dynamic transfer of state 'information' flowing between the casual
network 'gates' of the universe modeled as an evolving Big Bang Feynman Computer.

\section{Acknowledgments}

I am grateful to Anatoly A. Logunov, Sergei Klishevich, V. A. Petrov, and the
Organizing Committee of the XXIII International Workshop on the Fundamental
Problems of High Energy Physics and Field Theory, June 21-23, 2000, for their
generous invitation to share my ideas with them at the Institute for High
Energy Physics, Protvino, Moscow Region, Russia. Thanks to NSCL Staff; Roger
Zink, Dave Capelli, Reg Ronningen, Chris Ramsell, Doug Miller, James Sterling,
and the CCP/A1900 crew.

I would like to thank Toni Hitchcock and Pat Forrest and the rest of my
family. Thanks also to Lauren Eyres, Elaine Soller, E. Keith Hege, Gordon
Gilbert, James N. Lubbe, and Andre Bormanis.

\end{document}